\documentclass[
    aps,prb,twocolumn,
	groupedaddress,superscriptaddress,
	amsfonts,amssymb,amsmath,
	citeautoscript,longbibliography,
	letterpaper, nofootinbib
	]{revtex4-2}

\usepackage[utf8]{inputenc}
\usepackage[english]{babel}

\usepackage{microtype} 
\usepackage{xspace} 

\usepackage{txfonts}  
\usepackage{txfontsb} 

\usepackage{bm} 

\usepackage{xcolor}
\usepackage[]{graphicx} 
\graphicspath{{figs/}}

\usepackage[]{booktabs}
\usepackage{array}
\usepackage{layouts}
\usepackage{multirow}

\usepackage{enumerate}
\usepackage[inline]{enumitem}

\usepackage{xr}
\makeatletter
\newcommand*{\addFileDependency}[1]{
  \typeout{(#1)}
  \@addtofilelist{#1}
  \IfFileExists{#1}{}{\typeout{No file #1.}}
}
\makeatother

\usepackage{hyperref}
\hypersetup{colorlinks,
	linkcolor={blue!75!black!80!yellow},
	citecolor={blue!75!black!80!yellow},
	urlcolor={blue!75!black!80!yellow}
}

\usepackage[capitalize,nameinlink]{cleveref}

\crefname{subequations}{Eqs.}{Eqs.} 
\Crefname{subequations}{Eqs.}{Eqs.}
\crefformat{subequations}{#2Eqs.~(#1)#3}
\Crefformat{subequations}{#2Eqs.~(#1)#3}
\crefname{page}{p.}{p.} 
\crefname{table}{Table}{Tables}
\crefname{figure}{Figure}{Figures}
\crefname{section}{Section}{Sections}

\usepackage{placeins}

\usepackage{siunitx}
\DeclareSIUnit[number-unit-product = ]\percent{\char`\%} 

\usepackage[centering,hmargin=18mm,tmargin=29.4mm,bmargin=24mm]{geometry}

\usepackage{soul}

\usepackage{textcomp} 
\usepackage{xifthen}
\usepackage{etoolbox}
\newboolean{togglecomments}
\newboolean{toggletodos}
\newboolean{togglechanges}

\setboolean{togglecomments}{true}
\setboolean{toggletodos}{true}
\setboolean{togglechanges}{false} 

\newcommand{\textblacksquare}{$\blacksquare$}
\newcommand{\todo}[1]{\ifbool{toggletodos}%
	{\textcolor{green!60!black}{\small\textsf{{}\textsuperscript{\textsc{\textsf{todo}}}}[\ignorespaces#1]}} 
	{}}     
\newcommand{\comment}[2]{\ifbool{togglecomments}%
		{\textcolor{blue!70!black}{\small\sf\textsuperscript{\textsc{\textsf{\ignorespaces#1}}}[\ignorespaces#2]}} 
		{}}     

\newcommand{\reply}[2]{\ifbool{togglecomments}%
		{\textcolor{red!70!black}{\small\sf\textsuperscript{\textsc{\textsf{\ignorespaces#1}}}[\ignorespaces#2]}} 
		{}} 
        
\newcommand{\swap}[2]{\ifbool{togglechanges}
	{\ignorespaces#2}  
	{\textcolor{red!70!black}{[\ignorespaces#1]}\textrightarrow{}\textcolor{green!50!black}{[\ignorespaces#2]}}}
\newcommand{\remove}[1]{\ifbool{togglechanges}
	{}    
	{\textcolor{red!70!black}{\ignorespaces#1}}}
\newcommand{\inset}[1]{\ifbool{togglechanges}
	{\ignorespaces#1}  
	{\textcolor{green!50!black}{\ignorespaces#1}}}

\newcommand{\citeremind}[1]{%
	[\textcolor{blue!75!black!80!yellow}{\textblacksquare%
		\ifthenelse{\isempty{#1}}{}{\textsuperscript{\tiny\textsf{\ignorespaces#1}}}%
	}]\xspace}

\newcommand{\ie}{i.e.,\@\xspace} 

\newcommand{\im}{\mathrm{i}}

\newcommand{\appropto}{\mathrel{\vcenter{
			\offinterlineskip\halign{\hfil$##$\cr
				\propto\cr\noalign{\kern.2pt}\sim\cr\noalign{\kern-2.5pt}}}}}

\DeclareFontFamily{U}{mathx}{\hyphenchar\font45}
\DeclareFontShape{U}{mathx}{m}{n}{<5> <6> <7> <8> <9> <10>
                                  <10.95> <12> <14.4> <17.28> <20.74> <24.88>
                                  mathx10}{}
\DeclareSymbolFont{mathx}{U}{mathx}{m}{n}
\DeclareFontSubstitution{U}{mathx}{m}{n}

\makeatletter
\newcommand{\raisemath}[1]{\mathpalette{\raisem@th{#1}}}
\newcommand{\raisem@th}[3]{\raisebox{#1}{$#2#3$}}
\makeatother

\renewcommand{\paragraph}[1]{\vskip 1ex\noindent\textbf{#1.}~}

\usepackage{braket}
\usepackage[eulergreek]{sansmath}
\makeatletter
\renewcommand\@make@capt@title[2]{%
    \@ifx@empty\float@link{\@firstofone}{\expandafter\href\expandafter{\float@link}}%
    \sisetup{math-sf=\textsf}%
    \sansmath\sffamily\textbf{#1\@caption@fignum@sep}#2 
}%

\makeatother

\graphicspath{{figures/}}
\setboolean{togglecomments}{true}
\setboolean{toggletodos}{true}
\setboolean{togglechanges}{false} 
\usepackage[dvipsnames]{xcolor}

\begin{document}
\title{Zoology of chiral superconductors in Chern bands}

\author{André Grossi Fonseca}
\email{agfons@mit.edu}
\affiliation{Department of Physics, Massachusetts Institute of Technology, Cambridge, Massachusetts 02139, USA}
\affiliation{The NSF Institute for Artificial Intelligence and Fundamental Interactions}

\author{Aidan Reddy}
\affiliation{Department of Physics, Stanford University, Stanford, CA 94305, USA}
\affiliation{Department of Physics, Massachusetts Institute of Technology, Cambridge, Massachusetts 02139, USA}

\author{Ahmed Abouelkomsan}
\affiliation{Department of Physics, Massachusetts Institute of Technology, Cambridge, Massachusetts 02139, USA}

\author{Liang Fu}
\affiliation{Department of Physics, Massachusetts Institute of Technology, Cambridge, Massachusetts 02139, USA}

\author{Marin Solja\v ci\'c}
\affiliation{Department of Physics, Massachusetts Institute of Technology, Cambridge, Massachusetts 02139, USA}
\affiliation{The NSF Institute for Artificial Intelligence and Fundamental Interactions}
\affiliation{Research Laboratory of Electronics, Massachusetts Institute of Technology, Cambridge, Massachusetts 02139, USA}

\begin{abstract}
Evidence for chiral superconductivity has recently been observed in several van der Waals systems including rhombohedral multilayer graphene and twisted bilayer MoTe$_2$.
In the latter, superconductivity emerges at carrier densities near a fractional Chern insulator. This raises the question of what kinds of superconductors may emerge in a system of electrons in a topological band with strong repulsive interactions. 
Here, we adapt the target-phase optimization method to search for chiral superconductors in a minimal model of interacting electrons in a Chern band.
We construct a differentiable loss function for superconductors from the sign oscillation of the pair-binding energy and combine gradient-based optimization with a rigorous screening procedure to identify and characterize superconductors. Applying this framework within exact diagonalization to spinless electrons in periodically modulated Landau levels with screened Coulomb interactions, we uncover a broad family of chiral superconducting phases. At filling $\nu=2/3$, we recover the previously identified $f-\im f$ hole superconductors and find additional $p\pm \im p$ and $f+\im f$ superconductors of both electrons and holes, occurring near and far from the limit of ideal quantum geometry. At $\nu=1/2$, we identify $p- \im p$ electron and $f-\im f$ hole superconductors and find, for the first time, a direct transition between chiral superconductors and composite Fermi liquids. Our results reveal that chiral superconductivity in Chern bands comprises a diverse landscape of competing pairing instabilities and establish target-phase optimization as a general strategy for searching for quantum phases in complex interacting systems.
\end{abstract}
\maketitle 

\section{Introduction}

Moiré materials have emerged as a highly tunable platform for strongly correlated quantum matter. By combining narrow electronic bands, strong interactions, and 
tunable filling, these systems make it possible to engineer regimes in which electron correlations dominate over single-particle energy scales. This flexibility has led to the observation of a wide range of phenomena, including correlated insulators \cite{cao2018correlated}, unconventional superconductivity \cite{cao2018unconventional}, and orbital magnetism \cite{sharpe2019emergent}. 
In particular, the discovery of fractional quantum anomalous Hall states in such systems has opened 
a new setting for exploring strongly correlated phases in topological bands \cite{cai2023signatures, park_observation_2023, zeng2023thermodynamic, xu2023observation, lu_fractional_2024}. 

These observations have led to the natural question: are there phases of matter in these systems that go beyond those observed in partially filled Landau levels (LLs)? Recent exciting experiments have brought this question to the fore by reporting chiral superconductivity in the vicinity of fractional Chern insulators in rhombohedral graphene multilayers \cite{han2025signatures} and twisted MoTe$_2$ bilayers \cite{xu2025signatures}. 
Such observations establish a phenomenology that is impossible to access in 2D electron gases under a constant magnetic field, where continuous magnetic translation invariance disallows superconductivity \cite{guerci2026topological}.

Theoretically, an important connection has been established between zero-field moiré Chern bands and LLs subject to periodic magnetic fields \cite{tarnopolsky2019origin, morales2024magic}. 
This correspondence explains why such bands can realize fractional quantum Hall physics, while also providing a controlled route to study departures from the ``ideal'' LL limit \cite{ledwith2020fractional, wang2021exact}.
In particular, the Berry curvature and quantum metric of a general Chern band need not be uniform, and their momentum-space structure can qualitatively reshape the many-body ground state. Indeed, recent numerical and analytical work has shown that these geometric effects, together with bandwidth and particle-hole asymmetry~\cite{ahmed2020particle, ahmed2023quantum, hui2025ideal}, can drive a transition from a Laughlin-like fractional Chern insulator to an $f$-wave topological superconductor of holes near filling $\nu=2/3$, even in the presence of strong repulsive interactions \cite{guerci2025sc,guerci2026topological,wang2507chiral}.

Despite this progress, a broader understanding of superconductivity in topological bands remains lacking. Existing studies have focused primarily on $\nu=2/3$, motivated by recent experimental observations~\cite{han2025signatures, xu2025signatures}, and in bands with ideal quantum geometry. It is therefore unclear which distinct chiral superconducting orders can be stabilized in a generic Chern band, what are the nearby competing orders, and whether the superconductivity observed near $\nu=2/3$ is a special phenomenon or part of a larger family of chiral pairing instabilities. 

In this work, we adapt the recently developed target-phase optimization method \cite{fonseca2025gradient} to systematically search for chiral topological superconductors, hereafter referred to as chiral superconductors for brevity \cite{read2000paired}.
We combine gradient-based optimization of a target-phase loss function for superconductivity with a staged screening procedure designed to distinguish genuine paired phases from competing states and finite-size effects. We apply this approach to exact diagonalization of a model of modulated LLs under long-range repulsive interactions, which interpolates between an ideal LL limit and strongly non-ideal, inhomogeneous quantum geometry \cite{morales2024magic}. 
At $\nu=2/3$, we recover the $f-\im f$ hole superconductors reported in previous work \cite{guerci2025sc, wang2507chiral} and, in addition, identify novel $p \pm \im p$ and $f + \im f$ superconducting regimes both near and far from the ideal geometry limit, arising from pairing instabilities of both electron and hole Fermi liquids. 
We then extend the search to $\nu=1/2$, finding that this filling can also host chiral $p$-wave and $f$-wave superconductors, for both electron and hole carriers.  Upon tuning toward the ideal limit, we find for the first time numerical evidence for a direct phase transition from chiral superconductors into composite Fermi liquids, revealing a 
competition between chiral pairing and the most prominent compressible state in the lowest LL.
Due to the unit net magnetic flux per unit cell in our model, all of these superconductors are expected on general grounds to form vortex lattices with a vorticity of two per unit cell \cite{guerci2026topological}.

Our results establish that repulsive interactions among electrons in Chern bands can give rise to superconductors across a range of filling factors, with pairing symmetries strongly dependent on microscopic details.
The interplay of topology, quantum geometry, and interactions therefore goes beyond reproducing the hierarchy of fractional quantum Hall states, giving rise to a rich set of competing pairing orders.

\section{Target-phase optimization and screening for superconductivity}

We begin by describing our numerical strategy to search for chiral superconductivity based on the recently developed target-phase optimization method \cite{fonseca2025gradient}. The general philosophy is to turn the search for a quantum phase into an optimization problem. Rather than scanning a large Hamiltonian parameter space and applying a complete set of diagnostics at every point to identify the ground state, one first chooses a sharp feature of the desired phase and constructs a loss function from it, defined to be minimized in regions of parameter space where the chosen feature is most prominent. Gradient-based optimization is then used to minimize the loss function, thereby isolating parameter space regions where the feature is enhanced. The resulting candidate points are subsequently subjected to stringent post-processing checks, where other features of the phase are verified.

For superconductors, a simple feature that can be promoted to a loss function is the dependence of the ground-state energy on the parity of the number of electrons. 
In a finite-size conventional superconductor, charge sectors with even particle number are compatible with Cooper pairing, and therefore energetically favored relative to sectors containing an unpaired electron. This effect is traditionally quantified through the pair-binding energy
\begin{equation}
    E_{\rm B}(N_e)
    =
    E_0(N_e+2)+E_0(N_e)-2E_0(N_e+1),
    \label{eq:pair_binding}
\end{equation}
where $E_0(N_e)$ denotes the many-body ground-state energy in the charge sector with $N_e$ electrons. Conventional superconductors exhibit an ``even-odd effect'', \ie $E_{\rm B}(N_e)$ is negative for even $N_e$ and positive for odd $N_e$.
In contrast, in a weak-pairing spinless topological superconductor, the odd parity sector is favored due to an unpaired single-particle state at crystal momentum $\mathbf{K}=0$, leading to a sign-reversed ``odd-even effect" for periodic boundary conditions\footnote{Here we assume that the relevant single-particle dispersion has a single minimum at the $\Gamma$ point, and note that other minimum configurations can affect the even-odd effect.}\cite{read2000paired, fu2010teleportation, guerci2025sc}.

When the paired fermions are  doped holes on top of a filled band of electrons, the sign of the even-odd effect in $E_{\rm B}(N_e)$ depends on the number of available single-particle Bloch states $N_{\mathbf{k}}$, since the number of holes $N_h = N_{\mathbf{k}} - N_{e}$ depends on $N_{\mathbf{k}}$.
We use this fact, in addition to other many-body indicators, to distinguish electron and hole superconductors.

\begin{figure}
    \centering
    \includegraphics[scale=1]
    {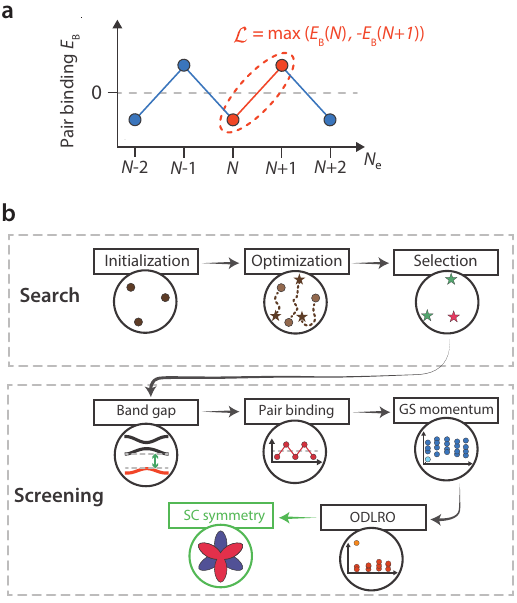}
    \caption{%
    Target-phase loss function for superconductivity and workflow.
    (a) Visual depiction of loss function definition in \cref{eq:sc_loss}, in terms of the sign oscillation of the pair binding energy, defined in \cref{eq:pair_binding}.
    (b) Schematic workflow for searching for and verifying superconductivity, through optimization of the superconductor loss function and a multi-step screening process (see text), respectively.
    }
    \label{fig:figure1}
\end{figure}

We now leverage this sign oscillation to construct a target-phase loss function for superconductors, which is defined in the space of Hamiltonian parameters $\{\mathbf{p} \}$.
Specifically, we define
\begin{equation}
    \mathcal{L}(\mathbf{p}, N_e)
    =
    \pm \max{(E_{\rm B}(\mathbf{p}, N_e), -E_{\rm B}(\mathbf{p}, N_e 
    +1))}
    \label{eq:sc_loss}
\end{equation}
where an overall plus/minus sign is used for superconductors exhibiting the even-odd/odd-even effect at the chosen system size and $N_e$.
By construction, $\mathcal{L}<0$ if and only if the pair-binding energy changes sign between the two consecutive charge sectors $N_e$ and $N_e + 1$. Thus, a negative loss indicates that the finite-size spectrum displays the sign oscillation expected of a paired state in those charge sectors, as schematically illustrated in \cref{fig:figure1}a. Therefore, we can optimize $\mathcal{L}$ in the parameter space, in search of regions with negative loss function.
In this work, we use exact diagonalization on small clusters to evaluate the pair binding energy, which has been recently employed to identify chiral superconductors in topological bands~\cite{guerci2025sc, guerci2026topological}.
We note that, at a higher computational cost, more charge sectors can in principle be included in the max function in \cref{eq:sc_loss} to more strongly enforce the sign oscillation of the pair binding energy across a broader range of fillings. 

We now summarize the search protocol (\cref{fig:figure1}b, top), and provide further details in the Supplemental Material (SM). The first step is to identify a Hamiltonian of interest (see Section III), for which we will search for chiral superconductors.
Then, we choose a finite cluster, a filling of interest, and whether we will target a hole or electron superconductor; these choices then fix $N_e$ and the overall sign of the loss function in \cref{eq:sc_loss}. In the Hamiltonian parameter space, we initialize several points on a coarse grid, which are optimized in parallel. For each initial value of $\mathbf{p}$, we use gradient-based optimization to minimize the loss, up until it attains a negative value, if at all. At each optimization step we diagonalize the Hamiltonian in the charge sectors entering \cref{eq:pair_binding} and \cref{eq:sc_loss} to evaluate the loss. Because each loss evaluation involves exponentially expensive exact diagonalization calculations, the search stage is carried out using small clusters to isolate potentially interesting regions of parameter space, after which more costly diagnostics at larger system sizes are applied.

Although a negative value of $\mathcal{L}$ is necessary to indicate pairing, it is not, by itself, sufficient to establish superconductivity. We therefore select all optimized points with $\mathcal{L} < 0$ and pass them through a staged screening pipeline (\cref{fig:figure1}b, bottom), which are several checks that the parameter points found must pass to be classified as chiral superconductors. Because they do not involve costly optimization, the screening steps are carried out on larger clusters than those used in the initial optimization, across multiple system sizes. First, because we work in a regime of band projection to the lowest band, we require that such band have a finite indirect band gap to the band above it, so that projecting to the lowest band is physically justified. Second, we check whether, for the larger cluster used in the pipeline, the pair binding sign is negative in the charge sector corresponding to the filling of interest. 
Third, we require the ground state to have vanishing total many-body crystal momentum, thereby restricting the present search to zero-momentum paired states.

The fourth and final stage is to check whether the state found exhibits off-diagonal long-range order (ODLRO). To do so, we first define the momentum-space two-body reduced density matrix (2RDM)
\begin{equation}
    \rho^{(2)}_{\mathbf{q}}(\mathbf{k},\mathbf{k}')
    =
    \left\langle
    c^\dagger_{\mathbf{k}}
    c^\dagger_{-\mathbf{k}+\mathbf{q}}
    c^{\phantom{}}_{-\mathbf{k}'+\mathbf{q}}
    c^{\phantom{}}_{\mathbf{k}'}
    \right\rangle .
    \label{eq:2rdm_fermions}
\end{equation}
where $c^{\phantom{}}_\mathbf{k}\, (c_\mathbf{k}^\dag)$ is the fermion annihilation (creation) operator at crystal momentum $\mathbf{k}$ and the expectation value is taken in the ground state. Due to momentum conservation, the 2RDM is block diagonal in different momentum transfer $\mathbf{q}$ sectors, each of which is an $N_s \times N_s$ matrix in indices $\mathbf{k}, \mathbf{k}'$, where $N_s$ is the number of unit cells.
We diagonalize each momentum block,
\begin{equation}
    \sum_{\mathbf{k}'}
    \rho^{(2)}_{\mathbf{q}}(\mathbf{k},\mathbf{k}')
    \chi^{(n)}_{\mathbf{q}}(\mathbf{k}')
    =
    \lambda^{(n)}_{\mathbf{q}}
    \chi^{(n)}_{\mathbf{q}}(\mathbf{k}),
    \label{eq:2rdm_eigenproblem}
\end{equation}
where $\lambda^{(n)}_{\mathbf{q}}$ are the pair-occupation eigenvalues and $\chi^{(n)}_{\mathbf{q}}(\mathbf{k})$ are the corresponding eigenvectors. A zero-momentum superconductor should exhibit its dominant two-particle correlation in the $\mathbf{q}=0$ sector. Then, following Yang's criterion for superconductivity~\cite{yang_longrange_1962}, the fourth step of the screening consists in checking whether the largest eigenvalue of the 2RDM $\lambda^{(N_s)}_{\mathbf{q}}$ occurs in the $\mathbf{q}=0$ block.

For the final surviving points, \ie those points which pass all stages of the screening, we extract the internal symmetry of the superconducting order from the condensate wavefunction, corresponding to the leading 2RDM eigenvector in the $\mathbf{q}=0$ sector, $\Psi_{\rm pair}(\mathbf{k}) \equiv \chi^{(N_s)}_{\mathbf{q}=0}(\mathbf{k})$. The phase winding of $\Psi_{\rm pair}(\mathbf{k})$ around the Fermi surface determines the angular momentum of the pair wave function; in particular, a winding by $\pm 2\pi$ corresponds to $p \pm \im p$ pairing, while a winding by $\pm 6\pi$ corresponds to $f \pm \im f$ pairing, which are the simplest possible symmetries for spinless fermions.

This staged workflow allows the target-phase loss to serve as an efficient proxy through parameter space, while the subsequent diagnostics enforce increasingly stringent criteria for genuine superconductivity. 
We use this procedure below to identify and classify chiral superconducting phases in a Hamiltonian describing modulated LLs.

\section{Microscopic model}

We now apply the optimization and screening procedures introduced above to a concrete model of topological bands. We choose a continuum Hamiltonian of spinless electrons in LLs subject to periodic modulations, which provides a controlled interpolation between the ideal lowest LL limit and Chern bands with nonuniform quantum geometry and finite dispersion. This construction is also motivated by twisted transition-metal dichalcogenide moir\'e systems, where the valence Chern bands can be understood in the ``adiabatic approximation'' as LLs in the presence of moir\'e-periodic scalar and magnetic fields~\cite{morales2024magic, guerci2025sc}. The model therefore captures, in a simplified setting, both the topology inherited from a parent LL and the geometric modulations induced by the moir\'e potential.

The single-particle Hamiltonian is
\begin{equation}
    H_0
    =
    \frac{1}{2m}
    \left[
    \mathbf{p}-e\mathbf{A}(\mathbf{r})
    \right]^2
    +
    V(\mathbf{r}),
    \label{eq:single_particle_model}
\end{equation}
where the magnetic field $B(\mathbf{r}) = \nabla \times \mathbf{A}(\mathbf{r}) \cdot \hat{z}$ is given by
\begin{equation}
    B(\mathbf{r})
    =
    B_0+\delta B(\mathbf{r}).
    \label{eq:modulated_field}
\end{equation}
Here $B_0$ is the uniform component 
while both $\delta B(\mathbf{r})$ and $V(\mathbf{r})$ 
have zero spatial average. We take these functions to have the translation symmetry of a triangular Bravais lattice enclosing one flux quantum per unit cell and retain only the lowest reciprocal-lattice harmonics:
\begin{align}
    \delta B(\mathbf{r})
    &=
    2\delta B_0
    \sum_{j=1}^{3}
    \cos\!\left(
    \mathbf{G}_j\cdot\mathbf{r}
    + \phi
    \right),
    \label{eq:magnetic_modulation}
    \\
    V(\mathbf{r})
    &=
    -2V_0
    \sum_{j=1}^{3}
    \cos\!\left(
    \mathbf{G}_j\cdot\mathbf{r}
    +
    \phi
    \right),
    \label{eq:scalar_modulation}
\end{align}
where $\mathbf{G}_1=\mathbf{b}_1, \mathbf{G}_2=\mathbf{b}_1+\mathbf{b}_2, \mathbf{G}_3=\mathbf{b}_2$ and $\mathbf{b}_{1,2}$ are the primitive reciprocal lattice vectors. The amplitudes $\delta B_0$ and $V_0$ tune the strength of the magnetic and scalar modulations, while $\phi$ controls their phase. 
Note that the Hamiltonian is invariant under the transformation $(\delta B_0, V_0, \phi) \rightarrow (-\delta B_0, -V_0, \phi + \pi).$

\begin{figure*}
    \centering
    \includegraphics[scale=1]
    {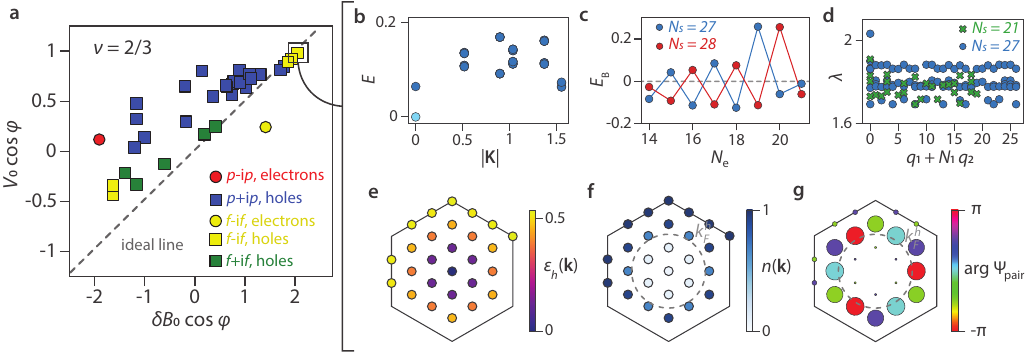}
    \caption{%
    Chiral superconductors at $\nu = 2/3$ found through target-phase optimization and screening steps.
    (a) 2D projection of parameter points found, with marker types denoting the charge carriers (circles/squares for electrons/holes) and color denoting the phase winding of the condensate wavefunction.
    Dashed line marks the ideal line $V_0 = \delta B_0/2$.
    Data in the panels are for parameter point highlighted, located at $(\delta B_0, V_0, \phi, U_0, d) = (2.06, 0.99, 0.07, 2.65, 1.37)$.
    All many-body data are calculated at system size $N_s = 27$ and number of electrons $N_e = 18$, unless otherwise specified.
    (b) Many-body spectrum as a function of the absolute value of the center-of-mass momentum $|\mathbf{K}|$.
    (c) Pair binding energy $E_{\rm B}$ as a function of number of electrons $N_e$.
    (d) Largest eigenvalues of the 2RDM $\lambda$ in each momentum transfer sector $\mathbf{q}$.
    (e) Interaction-induced hole dispersion $\varepsilon_h(\mathbf{k})$.
    (f) Ground-state momentum occupation $n(\mathbf{k})$, with non-interacting hole Fermi wavevector $k^h_F$ indicated, for $N_s = 21, N_e = 14$ and $N_s = 27, N_e = 18$.
    (g) Condensate wavefunction $\Psi_{\rm pair}$. The dot sizes denote the absolute value of the wavefunction, while the color denotes its phase.
    }
    \label{fig:figure2}
\end{figure*}

This model reduces to ordinary LLs when $\delta B_0 = V_0 = 0$. Furthermore, on the ``ideal line'' $V_0 = \delta B_0/2$ the lowest band is exactly flat and reproduces the ideal quantum geometry of the lowest LL, thereby realizing an Aharonov--Casher band~\cite{aharonovcasher}. 
In our calculations, we diagonalize \cref{eq:single_particle_model} in a LL basis and project the many-body problem into the lowest band, which for spinless electrons then implies that the number of Bloch states equals the number of unit cells, $N_\mathbf{k} = N_s$. We define the electron and hole filling to be $\nu = N_e/N_s, \nu_h = N_h/N_s$, respectively. The projected Hamiltonian is
\begin{equation}
    H
    =
    \sum_{\mathbf{k}}
    \varepsilon^{\phantom{}}_{\mathbf{k}}
    c^\dagger_{\mathbf{k}}c^{\phantom{}}_{\mathbf{k}}
    +
    \frac{1}{2A}
    \sum_{\mathbf{q}}
    U(\mathbf{q})
    :
    \bar{\rho}(\mathbf{q})
    \bar{\rho}(-\mathbf{q})
    : ,
    \label{eq:projected_hamiltonian_model}
\end{equation}
where $\varepsilon^{\phantom{}}_{\mathbf{k}}$ is the dispersion of the lowest band and $A$ is the system area. The projected density operator is $\bar{\rho}(\mathbf{q}) = \sum_{\mathbf{k}}\Lambda_{\mathbf{k}}(\mathbf{q}) c^\dagger_{\mathbf{k}+\mathbf{q}} c^{\phantom{}}_{\mathbf{k}}$, with form factor $\Lambda_{\mathbf{k}}(\mathbf{q}) = \langle u_{\mathbf{k}+\mathbf{q}}|u_{\mathbf{k}}\rangle$, where $|u_{\mathbf{k}}\rangle$ is the Bloch wave function of the projected band. The form factors encode the quantum geometry of the band and controls how the microscopic interaction acts after band projection.

Throughout this work we employ gate-screened Coulomb interactions
\begin{equation}
    U(\mathbf{q})
    =
     U_0 \frac{\tanh(|\mathbf{q}|d)}{|\mathbf{q}|d},
    \label{eq:gated_interaction_model}
\end{equation}
with $U_0$ parameterizing the interaction strength and $d$ the gate distance, which interpolates between a Coulomb, long-range regime at large $d$ to short-range interactions when $d \rightarrow 0$.
Therefore, the parameter space of this model is five-dimensional, $\mathbf{p} = (\delta B_0, V_0, \phi, U_0, d)$.
Throughout, we work in a positive charge background, which amounts to omitting the $\mathbf{q}=0$ component of the interaction. Additionally, we define all energies in units of $\hbar\omega_c = \hbar e B_0/m$, all magnetic field strengths in units of $B_0$, and all lengths in units of the magnetic length $\ell_B = \sqrt{\hbar/e B_0}$, which for one flux quantum per unit cell implies unit cell area $A_{\operatorname{UC}} = 2\pi$.

Previous work has studied variants of this model but restricted to the limit of ideal quantum geometry, finding chiral $f$-wave superconductivity for strong enough periodic modulations~\cite{guerci2026topological, guerci2025sc, wang2507chiral}. Our setting generalizes these studies and provides a broad setting to investigate the emergence of chiral superconductivity in Chern bands beyond the ideal limit with nonuniform quantum geometry and finite energy dispersion.

\section{Chiral superconductivity at $\nu=2/3$}

We first apply our search procedure at filling $\nu=2/3$, for which there is experimental evidence of chiral superconductivity in the vicinity of fractional Chern insulators in moir\'e materials \cite{xu2025signatures}. Furthermore, recent numerical work on closely related models of modulated LLs identified an $f - \im f$ superconductor of holes emerging near $\nu=2/3$ \cite{guerci2026topological, guerci2025sc,wang2507chiral}. This filling therefore provides both an experimentally motivated target and a useful benchmark: a successful optimization procedure should be able to recover the previously reported chiral $f$-wave phase, while also revealing whether other superconducting orders appear away from the parameter regimes studied before.

For the optimization stage, we use a cluster with $N_s = 16$ unit cells, which has a manageable size for gradient-based optimization, while still preserving $C_3$ symmetry about its center (see SM for the exact diagonalization clusters used in this work). We compute the superconductor loss function at $N_e = 11$, which approximates the filling of interest on this cluster. Since the number of single-particle orbitals is even, electron and hole chiral superconductors should exhibit the same odd-even effect for this system size. Therefore, we use the same positive overall sign in \cref{eq:sc_loss}, and distinguish the different superconductors during the screening stages, where both the sign of the pair-binding energy and the ground-state momentum occupation determine the charge carriers. Candidate points are then screened on 
larger clusters with $N_s = 27$ and $N_s = 28$. For electron superconductivity, we employ electron number $N_e = 17$ and $N_e = 19$ respectively; for hole superconductivity we use $N_e = 18$ and $N_e = 19$ on the same clusters, which then amounts to an odd number of holes $N_h = 9$ in both system sizes (see SM for further details).

\cref{fig:figure2}a summarizes the rich zoology of chiral superconductors we have found. Firstly, note that, because of the symmetry $(\delta B_0, V_0, \phi) \rightarrow (-\delta B_0, -V_0, \phi + \pi)$, and the fact that most points found have phase $\phi$ near 0 or $\pi$ (see SM), we find it more meaningful to plot the points found in $(\delta B_0 \cos \phi, V_0 \cos \phi)$ space. 
As a benchmark, the optimization reliably rediscovers the $f-\im f$ hole superconductors found in previous work along the ideal line at strong modulation strength (yellow squares), albeit now in the presence of long-range interactions. We find that this phase is not confined to the ideal line, with a second pocket of $f-\im f$ hole superconductors found in a region with opposite modulation sign $\delta B_0,  V_0 < 0$, corresponding to a partially filled non-ideal, dispersive band. 
For weaker modulations and both near and substantially away from ideality, we find novel $p  +\im p$ and $f + \im f$ hole superconductors, as well as $p  -\im p$ and $f - \im f$ electron superconductors. 
We note that, because \cref{fig:figure2}a is a 2D projection, superconducting regions that appear close together here may be far apart in the full five-dimensional parameter space, and therefore proximity does not imply the existence of direct phase transitions between different superconductors.
 
We highlight the evidence for superconductivity for a representative point in the $f - \im f$ hole region in \cref{fig:figure2}(b-g), and show analogous data for the other types of superconductors in the SM.
The exact diagonalization spectrum has a single ground state at total momentum $\mathbf{K}=0$ (\cref{fig:figure2}b), consistent with zero-momentum pairing. The pair-binding energy exhibits the expected sign oscillation across a broad range of filling factors including $\nu=2/3$, 
with the oscillation pattern flipping sign for two consecutive system sizes, consistent with chiral superconductivity of hole carriers (\cref{fig:figure2}c). The 2RDM further corroborates superconductivity: its leading eigenvalue lies in the $\mathbf{q}=0$ momentum transfer sector and increases with system size at fixed filling, consistent with the development of ODLRO (\cref{fig:figure2}d).

The microscopic structure of this state is also consistent with weak-pairing hole superconductivity. We compute the effective hole dispersion of the lowest band by combining the electron band dispersion with the Hartree--Fock self-energy generated by the filled band~\cite{guerci2025sc}. The resulting hole band has a single minimum at the $\Gamma$ point (\cref{fig:figure2}e), which leads to a simple hole Fermi surface centered at $\Gamma$, as we confirm through the ground-state momentum occupation $n(\mathbf{k}) = \langle c^\dagger_{\mathbf{k}}c^{\phantom{}}_{\mathbf{k}} \rangle$, which shows a sharp drop across a momentum shell centered at the origin, which coincides with the non-interacting Fermi wavevector for holes, $\mathbf{k}^h_F = \sqrt{4 \pi \nu_h/A_{\operatorname{UC}}} = \sqrt{2 \nu_h}$ (\cref{fig:figure2}f). Finally, the leading $\mathbf{q}=0$ eigenvector of the 2RDM $\Psi_{\rm pair}(\mathbf{k})$ peaks near $\mathbf{k}_F$, consistent with weak-pairing topological superconductivity, with its phase winding by $-6\pi$ around $\Gamma$ (\cref{fig:figure2}g).
These diagnostics all unambiguously point to topological superconductivity with chiral $f-\im f$ pairing, as found in previous work.

\begin{figure*}
    \centering
    \includegraphics[scale=1]
    {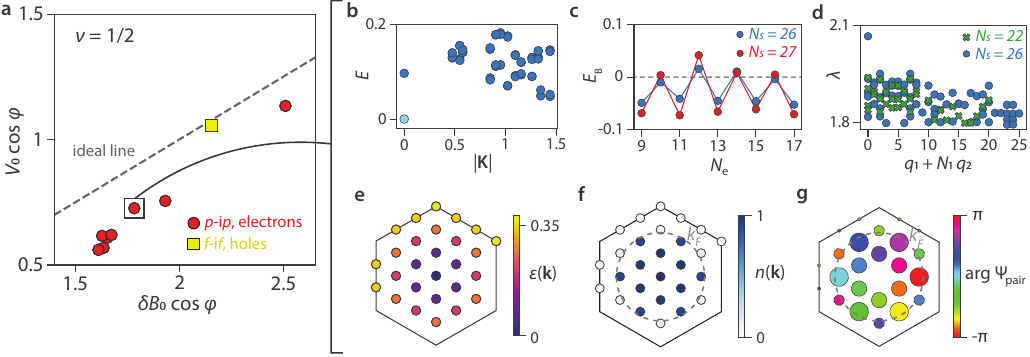}
    \caption{%
    Chiral superconductors at $\nu = 1/2$ found through target-phase optimization and screening steps.
    (a) 2D projection of parameter points found, with marker types denoting the charge carriers (circles/squares for electrons/holes) and color denoting the phase winding of the condensate wavefunction.
    Dashed line marks the ideal line $V_0 = \delta B_0/2$.
    Data in the panels are for the parameter point highlighted, located at $(\delta B_0, V_0, \phi, U_0, d) = (-1.79, -0.73, 3.18, 1.98, 0.77)$.
    All many-body data are calculated at system size $N_s = 26$ and number of electrons $N_e = 13$, unless otherwise specified.
    (b) Many-body spectrum as a function of the absolute value of the center-of-mass momentum $|\mathbf{K}|$.
    (c) Pair binding energy $E_{\rm B}$ as a function of number of electrons $N_e$.
    (d) Largest eigenvalues of the 2RDM $\lambda$ in each momentum transfer sector $\mathbf{q}$ for $N_s = 22, N_e = 11$ and $N_s = 26, N_e = 13$.
    (e) Bare band dispersion $\varepsilon(\mathbf{k})$.
    (f) Ground-state momentum occupation $n(\mathbf{k})$ for $N_s = 27, N_e = 13$, with non-interacting electron Fermi wavevector $k_F$ indicated.
    (g) Condensate wavefunction $\Psi_{\rm pair}$ for $N_s = 27, N_e = 13$. The dot sizes denote the absolute value of the wavefunction, while the color denotes its phase.
    }
    \label{fig:figure3}
\end{figure*}

At the single-particle level, although the $f-\im f$ superconductor analyzed sits at a point with strong modulations, these are not enough to lead to a band inversion of the lowest Chern band, and we find Chern number $C=+1$, matching that of the lowest LL. 
Furthermore, this point is very close to the ideal line, suggesting that the fractionally filled band is nearly ideal.
This can be quantitatively assessed through the quantum geometry of the band, specifically through the ``trace violation'' $T$ and Berry curvature fluctuations $\sigma_\Omega$ (see SM for definitions).
The property $T = 0$ defines an ideal band and holds throughout the ideal line. For this point we find $\sigma_\Omega \approx 0.46, T \approx 0.11$, indicating a nearly ideal band but with considerable Berry curvature fluctuations.

Taken together, the $\nu=2/3$ results show that the target-phase search not only reproduces the chiral $f$-wave hole superconductor, but also uncovers a richer set of competing chiral superconducting orders enabled by departures from lowest-LL quantum geometry.
This establishes that the previously identified $f$-wave state is part of a broader and richer landscape of superconductivity at $\nu=2/3$.

\section{Chiral superconductivity at $\nu=1/2$}

We next turn to filling $\nu=1/2$, which both provides a test of the generality of our method across different fillings and access to distinct quantum phases. 
In the half-filled lowest LL, Coulomb interactions are expected to favor a compressible composite Fermi liquid rather than a gapped fractional quantum Hall state. 
It is therefore not evident a priori whether departures from ideal LL band geometry can stabilize superconductivity at this filling, or which pairing channels may emerge.

We once again split the method into an optimization and screening stage. 
For the first part, we employ the same 16-site cluster as before, but now compute the superconductor loss function at $N_e = 8$. Once more, we use the positive sign in \cref{eq:sc_loss} to target weak-pairing, topological superconductors, and distinguish hole and electron superconductors during screening. For the second part, we use clusters with $26$ and $27$ sites. For electron superconductivity, we use electron number $N_e = 13$ for both system sizes, as this is the charge sector closest to half filling for the 27-site cluster; for hole superconductivity, we use $N_e = 13$ and $N_e = 14$ on the same clusters, respectively.

\cref{fig:figure3}a summarizes the superconducting phases found at $\nu=1/2$. In contrast to the larger variety of orders identified at $\nu=2/3$, our search finds two principal phases: a $p-\im p$ superconductor of electrons and an $f-\im f$ superconductor of holes. The $p-\im p$ phase seems to be favored away from the ideal line, while the $f-\im f$ superconductor appears near the ideal line, but at large LL modulations. 
Notably, the $f-\im f$ superconductor found here seems to be the same superconductor as the one found in the $\nu=2/3$ search but now near $\nu=1/2$, which is consistent with previous findings that stronger LL modulations stabilize superconductivity down to lower fillings~\cite{guerci2026topological}.
For a representative point in the $p - \im p$ region (see SM for analogous data for the $f - \im f$ superconductor), we show the same diagnostics employed at $\nu=2/3$, which provide strong evidence for chiral superconductivity. The ground state occurs at total momentum $\mathbf{K}=0$ (\cref{fig:figure3}b); the pair-binding energy displays the expected sign oscillation across neighboring charge sectors, with the same sign structure for even and odd system sizes, consistent with a superconductor of electron carriers (\cref{fig:figure3}c); further, the dominant eigenvalue of the 2RDM lies in the $\mathbf{q}=0$ sector and grows for larger system sizes, signaling ODLRO. 

Because we now focus on an electron superconductor, the relevant single-particle dispersion is the bare lowest electron band, which we show in \cref{fig:figure3}e.
It once again has a minimum at $\Gamma$, with the momentum occupation centered around it and showing a clear sign of an electron Fermi surface, with occupation drop near the non-interacting Fermi wavevector for electrons, $\mathbf{k}_F = \sqrt{2 \nu}$  (\cref{fig:figure3}f).
Finally, the leading condensate wavefunction is once again concentrated near the Fermi surface, and its phase displays a $-2\pi$ winding, characterizing the paired state as a $p-\im p$ superconductor.
At the single-particle level, we once again find Chern number $C = +1$, with trace violation and Berry curvature fluctuations $T \approx 0.79, \sigma_\Omega \approx 0.16$, indicating a band far from the ideal limit.

The $\nu=1/2$ results broaden the scope of our search. They show that the optimization does not merely rediscover superconductivity near the previously studied $\nu=2/3$ regime, but can identify distinct electron- and hole-paired phases at any filling. More generally, the success of the same target-phase loss and screening pipeline across both fillings demonstrates that the method provides a flexible and systematic framework for discovering superconducting order in multi-parameter many-body systems.

\section{SC-to-CFL phase transition}

Our results at $\nu = 1/2$ demonstrate that chiral superconductivity in topological bands is not restricted to fillings proximate to fractional Chern insulators, but can also arise at a filling near compressible ground states.
These superconductors are also notable because they occur relatively close to the ideal line, where we expect the physics to be adiabatically connected to the half-filled lowest LL for low to moderate modulations. Therefore, tuning the modulations of these superconductors toward the low-modulation ideal limit could lead to a direct phase transition from chiral superconductors into composite Fermi liquids.

\begin{figure}
    \centering
    \includegraphics[scale=1]
    {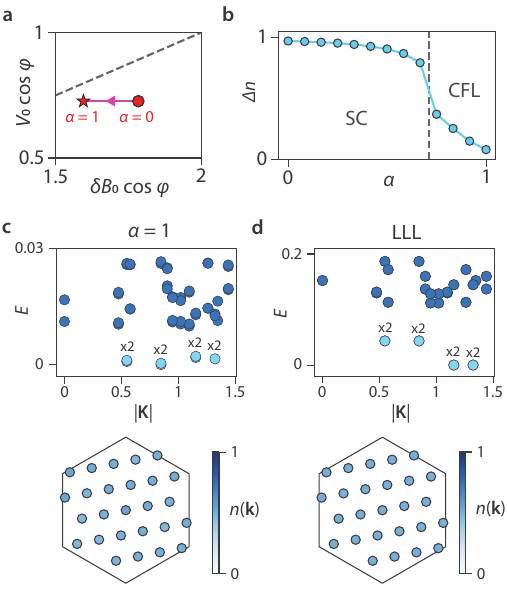}
    \caption{%
    Quantum phase transition between a $p - \im p$ superconductor (SC) and a composite Fermi liquid (CFL). All data are calculated at system size $N_s = 26$ and number of electrons $N_e = 13$.
    (a) Parameter space path taken to drive phase transition, which corresponds to sweeping the path $\delta B_0 (\alpha) = (1-\alpha) \delta B^0_0  + \alpha \delta B^f_0$, with $\delta B_0^0 = -1.79$, $\delta B_0^f = -1.60$, while holding all other parameters fixed to $(V_0, \phi, U_0, d) = (-0.73, 3.18, 1.98, 0.77)$.
    (b) Ground-state momentum occupation range $\Delta n = \max n(\mathbf{k}) - \min n(\mathbf{k})$ as a function of sweep parameter $\alpha$.
    (c) Many-body spectrum (top) and ground-state momentum occupation (bottom) for final point in the sweep $\alpha=1$.
    (d) Many-body spectrum (top) and ground-state momentum occupation (bottom) for the lowest Landau level at $\nu=1/2$ and short-range interactions.
    }
    \label{fig:figure4}
\end{figure}

We find that this is indeed the case for the $p - \im p$ superconductor.
\cref{fig:figure4}a shows the path in parameter space we take, which starts at a chiral superconductor and moves towards the ideal line. \cref{fig:figure4}b shows the ground-state momentum occupation range $\Delta n = \max n(\mathbf{k}) - \min n(\mathbf{k})$ throughout the path. Its value starts out near unity, due to the Fermi surface associated with the superconductor (see \cref{fig:figure3}f). However, eventually $\Delta n$ sharply drops towards zero, indicating a phase transition to a phase with no conventional Fermi surface.
\cref{fig:figure4}c, top shows the many-body spectrum at the final point of the parameter path, whose pattern of low-energy states and degeneracies bears close resemblance to the low-energy spectrum of the half-filled lowest LL under short-range interactions $U(\mathbf{q}) = -q^2$ ( \cref{fig:figure4}d, top). And indeed,  the momentum occupation of both the point found (\cref{fig:figure4}c, bottom) and the half-filled lowest LL (\cref{fig:figure4}d, bottom) are fairly flat, ruling out a conventional Fermi liquid. These features provide, for the first time, strong numerical evidence for a direct transition between chiral superconductors and composite Fermi liquids.
Such a phase transition is likely first-order, as a continuous transition between these phases must be fine-tuned~\cite{yunchao2026cfl}.
We have found that a similar tuning from the $f - \im f$ superconductor towards the low-modulation ideal regime seems to lead to an intervening Fermi liquid phase.

\section{Conclusion}
In this work, we introduced a systematic numerical method for efficiently discovering chiral superconductivity in general many-body systems. Our approach adapts the target-phase optimization scheme~\cite{fonseca2025gradient} to superconductors by constructing a loss function from the even-odd effect of the pair-binding energy. Candidate points found through optimization of the loss function are then subjected to a rigorous staged screening procedure that verifies superconductivity. 
Applying this method to a model of periodically modulated LLs relevant for moiré materials, we uncovered a zoology of chiral superconductors at fillings $\nu=2/3$ and $\nu=1/2$. 
We both recover the previously reported $f-\im f$ hole superconductor at $\nu = 2/3$ in a band with ideal quantum geometry, and also greatly expand on it, finding $p \pm \im p$ and $f + \im f$ superconductors of electrons and holes, all of which form vortex lattices with doubled vorticity per unit cell \cite{guerci2026topological}.
Our results reveal a wealth of superconducting orders that can emerge at different fillings of Chern bands, while emphasizing that the specific pairing symmetries are largely detail-dependent.
Finally, we find a novel phase transition between a $p - \im p$ superconductor and a composite Fermi liquid, greatly enriching the landscape of quantum phases and transitions in moiré systems~\cite{yunchao2026cfl}.

Several directions naturally follow from our results. 
A better understanding of the mechanisms behind the different pairings found here is an important direction for future work.
Applying this framework to more realistic models, especially continuum moiré Hamiltonians, would be desirable, while larger-scale calculations using density-matrix renormalization group, variational Monte Carlo, or related many-body methods could establish the thermodynamic stability of the phases identified. Extending the target-phase loss function to finite-momentum pairing would make it possible to search for pair-density waves, while incorporating spin, valley, and multiple active bands could lead to the discovery of novel multi-component superconductors. 
In addition, our method provides a systematic way to search for microscopic models hosting ``anyon superconductors'', obtained by doping fractional Chern insulators ~\cite{Laughlin1988Superconducting,Shi2024Doping, Divic2024Anyon, Nosov2025Anyon, Pichler2025Microscopic}.
More broadly, our results show that target-phase optimization provides a general and scalable strategy for exploring ordered phases in complex interacting systems, and that topological bands host a diverse landscape of chiral superconductors, which directly compete with the fractional quantum Hall states in the lowest LL.

\section{Acknowledgments}
A.G.F. and M.S. acknowledge support from the National Science Foundation under Cooperative Agreement PHY-2019786 (The NSF AI Institute for Artificial Intelligence and Fundamental Interactions).
M.S.\ acknowledge support from the U.S.\ Office of Naval Research (ONR) Multidisciplinary University Research Initiative (MURI) under Grant No.\ N00014-20-1-2325 on Robust Photonic Materials with Higher-Order Topological Protection.
This material is based upon work also supported in part by the U. S. Army Research Office through the Institute for Soldier Nanotechnologies at MIT, under Collaborative Agreement Number W911NF-23-2-0121. A.P.R. acknowledges support from the GLAM Postdoctoral Fellowship at Stanford University.
The MIT SuperCloud and Lincoln Laboratory Supercomputing Center provided computing resources that contributed to the results reported in this work. 
Any use of generative AI in this manuscript adheres to ethical guidelines for use and acknowledgment of generative AI in academic research. 
Each author has made a substantial contribution to the work, which has been thoroughly vetted for accuracy, and assumes responsibility for the integrity of their contributions~\cite{mann2024ai}.

\bibliography{references}
\end{document}


\title{\texorpdfstring{
        SUPPLEMENTAL MATERIAL\\[1ex]
        Zoology of chiral superconductors in Chern bands
        }
        {Supplemental Material}
       }

\author{André Grossi Fonseca}
\email{agfons@mit.edu}
\affiliation{Department of Physics, Massachusetts Institute of Technology, Cambridge, Massachusetts 02139, USA}
\affiliation{The NSF Institute for Artificial Intelligence and Fundamental Interactions}

\author{Aidan Reddy}
\affiliation{Department of Physics, Stanford University, Stanford, CA 94305, USA}
\affiliation{Department of Physics, Massachusetts Institute of Technology, Cambridge, Massachusetts 02139, USA}

\author{Ahmed Abouelkomsan}
\affiliation{Department of Physics, Massachusetts Institute of Technology, Cambridge, Massachusetts 02139, USA}

\author{Liang Fu}
\affiliation{Department of Physics, Massachusetts Institute of Technology, Cambridge, Massachusetts 02139, USA}

\author{Marin Solja\v ci\'c}
\affiliation{Department of Physics, Massachusetts Institute of Technology, Cambridge, Massachusetts 02139, USA}
\affiliation{The NSF Institute for Artificial Intelligence and Fundamental Interactions}
\affiliation{Research Laboratory of Electronics, Massachusetts Institute of Technology, Cambridge, Massachusetts 02139, USA}

\maketitle

\setlength{\parindent}{0em}
\setlength{\parskip}{.5em}

\noindent{\small\textbf{\textsf{CONTENTS}}}\\ 
\twocolumngrid
\begingroup
    \let\bfseries\relax 
    \deactivateaddvspace 
    \deactivatetocsubsections 
    \makeatletter\@starttoc{toc}\makeatother 
\endgroup
\onecolumngrid

\count\footins = 1000 
\interfootnotelinepenalty=10000 

\section{Details of target-phase optimization}
\label{sec:optim_details}

\begin{figure}
    \centering
    \includegraphics[scale=1]
    {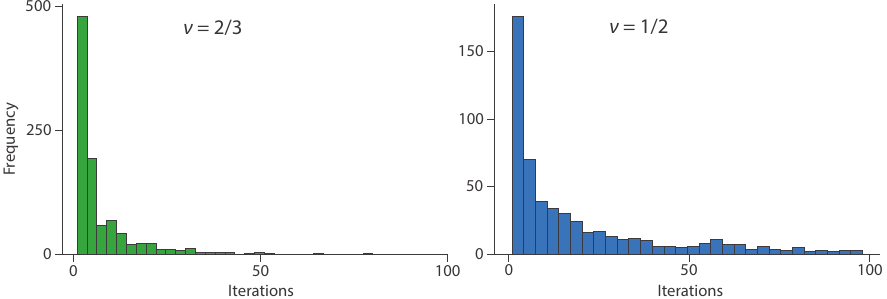}
    \caption{%
    Histograms for the number of Sobol grid points that reached points with negative loss, with the number of iterations on the $x$ axis. Data are shown for optimizations done at fillings $\nu = 2/3$ (left) and $\nu = 1/2$ (right).
    }
    \label{fig:hist}
\end{figure}

Here we provide further details on the optimization of the superconductor loss function $\mathcal{L}$.
Before setting up the workflow, we first carry out a reparametrization of the parameter space, to new variables over which $\mathcal{L}$ can be more conveniently optimized.
Specifically, we first note that the line $U_0 = 0$ consists of free electrons, and therefore the pair-binding energy should generically vanish (modulo shell filling effects), which likely leads to a local minimum that traps the optimization algorithm. Therefore, we first consider optimizing over its inverse $U_0^{-1}$, for which the free-fermion limit is sent to infinity. Additionally, because we are interested in the regime of repulsive interactions, we must avoid negative values of $U_0$. Therefore, we reparametrize $U_0^{-1} \equiv \operatorname{softplus}( z_U)$, where $\operatorname{softplus}(x) \equiv \log (1+e^x)$ is always non-negative.
We would also like to restrict to physical positive values of the gate distance $d$; furthermore, the regimes $d \ll 1$ and $d \gg 1$ correspond to qualitatively distinct interactions, so they must be considered on equal footing, even though the small-$d$ regime is much smaller in parameter space than its counterpart. To accommodate these two aspects, we reparametrize $d \equiv \sinh (\operatorname{softplus}(z_d))$ so that the optimizer can more finely  explore the small-$d$ region, while navigating the large-$d$ region more coarsely.
Therefore, the parameter space that we optimize over is $(\delta B_0, V_0, \phi, z_U, z_d)$, and at the end we convert $(z_U, z_d)$ back to the physical variables $(U_0, d)$.

\begin{figure}
    \centering
    \includegraphics[scale=1]
    {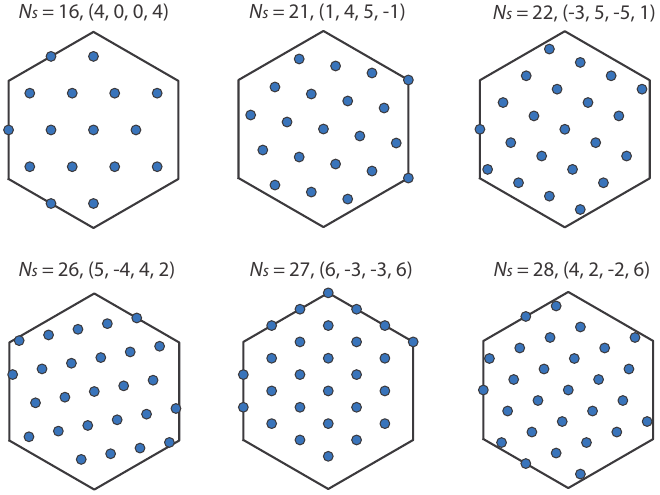}
    \caption{%
    Momentum meshes for the finite clusters used in this work. $N_s$ denotes the number of unit cells, and corresponding cluster integers are written as ($n_{x1}, n_{y1}, n_{x2}, n_{y2}$).
    }
    \label{fig:figure1}
\end{figure}

The first step in the workflow is to initialize several parameter seeds on a coarse grid, which are then optimized independently. In this work, we use the same grid for $\nu=2/3$ and $\nu=1/2$, which is a Sobol grid of 1000 initial points, initialized within the window $\delta B_0 \in [-2, 2], V_0 \in [0, 2], \phi \in [0, 2\pi], U_0^{-1} \in [0.1, 2.0], d \in [0.01, 3.0]$.
We have exploited the Hamiltonian symmetry $\phi \rightarrow \phi + \pi, \delta B_0 \rightarrow -\delta B_0, V_0 \rightarrow -V_0$ to restrict the $V_0$ window to non-negative values.
Note that, apart from the constraints enforced by the reparameterization described above, the parameter points are free to move anywhere in parameter space during the optimization, not being restricted to remain within the initialization region.
However, points hosting superconductors that sit far from this window are less likely to be found by the optimization.

We employ the Adam optimizer for faster convergence, with gradients computed by automatic differentiation through the exact diagonalization calculation, which must be carried out at each step of the optimization. We use a learning rate of $0.1$, and terminate the optimization as soon as $\mathcal{L} < 0$ or when the number of iterations reaches 100.

\cref{fig:hist} shows histograms for the number of iterations required to reach negative loss, for those parameter initializations which did find negative loss points. Out of the $1000$ Sobol initializations, 992 runs successfully terminated at points with $\mathcal{L} < 0$ at $\nu=2/3$, whereas 543 did so at $\nu=1/2$, consistent with our later findings that there are more superconductors at $\nu=2/3$ than at half filling. At both fillings the optimizer quickly finds potential superconductors, with more than half of the runs converging with fewer than 6 iterations at $\nu=2/3$, and fewer than 11 iterations at $\nu=1/2$. This efficiency underscores the power and utility of the target-phase optimization method for exploring high-dimensional parameter spaces. 

\section{Exact diagonalization clusters used}

In order to confirm the superconducting nature of the ground states we studied in the main text, several different finite clusters had to be employed.
These are constructed as supercells of the triangular lattice with primitive lattice vectors $ \mathbf{a}_{1,2}  = (\pi/\sqrt{3})^{1/2}(\sqrt{3}, \mp 1)$, defined such that the unit cell area is $2\pi$. The supercell boundary vectors are taken to be linear combinations of the lattice vectors, \ie $\mathbf{L}_1 = n_{x1} \mathbf{a}_1 + n_{y1} \mathbf{a}_2,\, \mathbf{L}_2 = n_{x2} \mathbf{a}_1 + n_{y2} \mathbf{a}_2$ for cluster integers $n_{x1}, n_{y1}, n_{x2}, n_{y2}$. This allows for greater control over the aspect ratio, high-symmetry momenta sampled and point group symmetry of the cluster for a given system size. For example, the clusters used for $N_s = 21, 27, 28$ have exact $C_3$ rotational symmetry about the $\Gamma$ point, with the first two containing the $K$ and $K'$ points.
For a cluster with $N_s$ unit cells, the four cluster integers must satisfy $|n_{x1} n_{y2} - n_{x2}n_{y1}| = N_s$ (see \cite{repellin2014z2} for more details).
\cref{fig:figure1} shows the momentum meshes for each of the clusters employed in the main text.

\section{Superconductors in parameter space}

\begin{figure}
    \centering
    \includegraphics[scale=1]
    {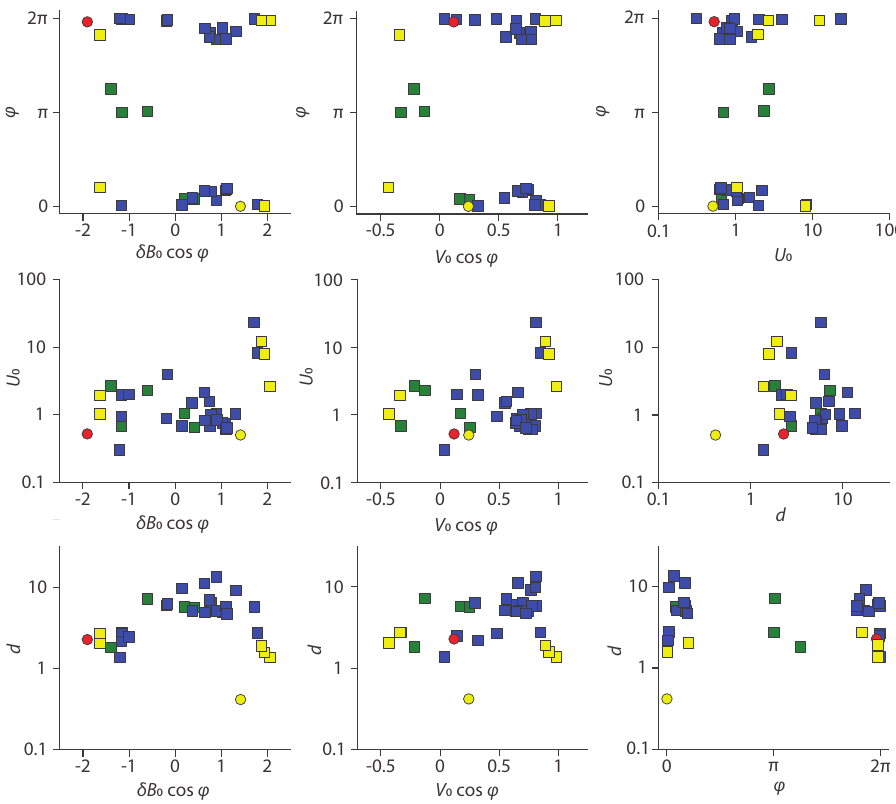}
    \caption{%
    Chiral superconductors at $\nu = 2/3$, projected along different parameter directions.
    }
    \label{fig:figure2}
\end{figure}

\begin{figure}
    \centering
    \includegraphics[scale=1]
    {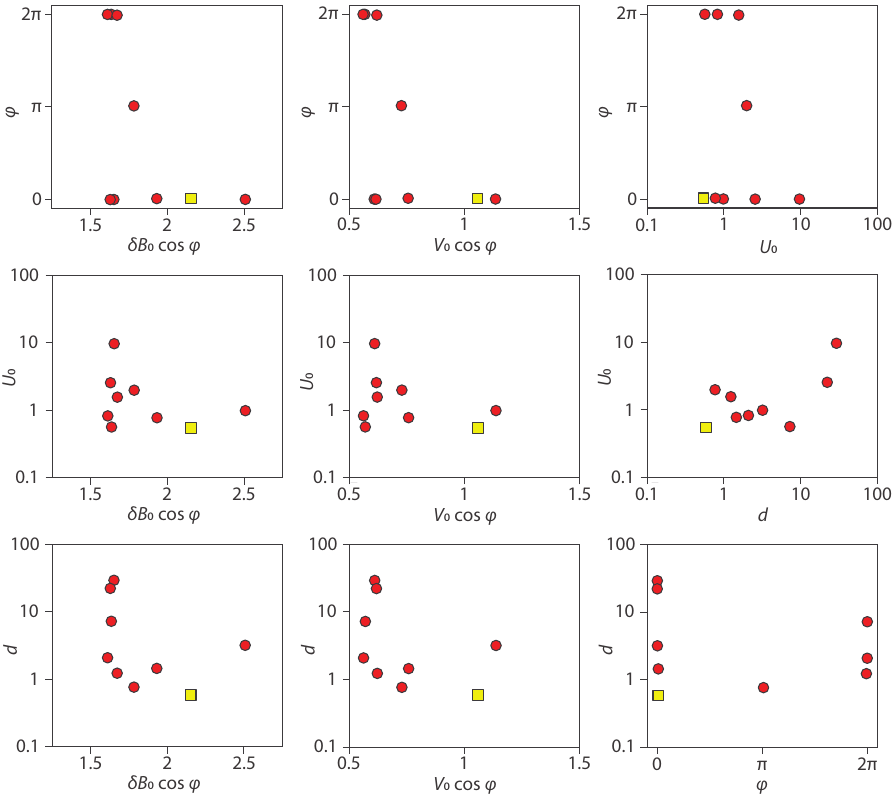}
    \caption{%
    Chiral superconductors at $\nu = 1/2$, projected along different parameter directions.
    }
    \label{fig:figure2b}
\end{figure}

In \cref{fig:figure2} and \cref{fig:figure2b} we show the same chiral superconductors as shown in the main text, but now on different 2D projections in parameter space. One notable feature is that most points found seem to lie at relatively small modulation angles $\phi$ away from $\phi=0, \pi$, a finite value of which breaks inversion symmetry.
This is naturally explained by the observation that inversion symmetry typically favors superconductivity by enforcing degenerate single-particle orbitals at $\mathbf{k}$ and $-\mathbf{k}$, as time-reversal symmetry is broken; here, we provide numerical evidence for this from a purely data-driven analysis.

\section{Numerical evidence for other chiral superconductors}

\begin{figure}
    \centering
    \includegraphics[scale=1]
    {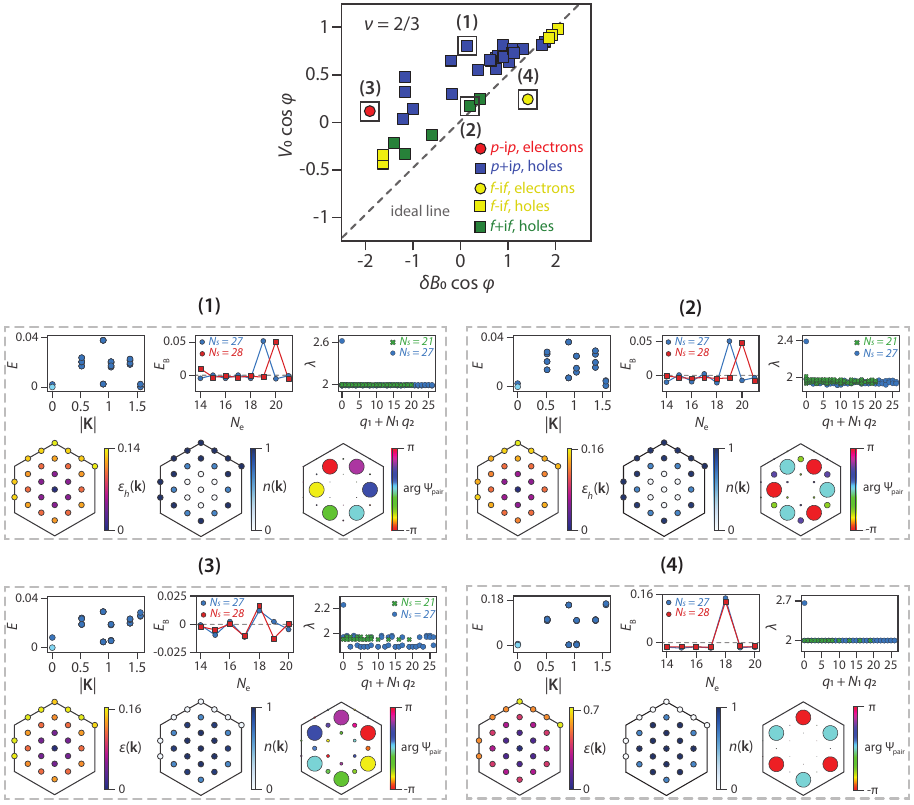}
    \caption{Evidence for chiral superconductivity for other representative points at $\nu = 2/3$.
    Their locations in parameter space $(\delta B_0, V_0, \phi, U_0, d)$ are respectively: (1)  $(0.15, 0.81, 0.06, 0.69, 9.80)$; (2)  $(0.21, 0.18, 0.26, 1.06, 5.78)$; (3) $(-1.92, 0.12, 6.17, 0.52, 2.28)$; (4) $(1.42, 0.24, 0.01, 0.51, 0.42)$.
    Calculations for hole superconductors are carried out at the same system sizes and electron numbers as the main text.
    For electron superconductors, many-body spectra, momentum occupation and pair condensate wavefunction were computed at system size $N_s = 27$ and number of electrons $N_e = 17$, whereas the 2RDM spectra were computed at $N_s = 21, N_e = 13$ and $N_s = 27, N_e = 17$. 
    }
    \label{fig:other_scs_two_thirds}
\end{figure}

\begin{figure}
    \centering
    \includegraphics[scale=1]  {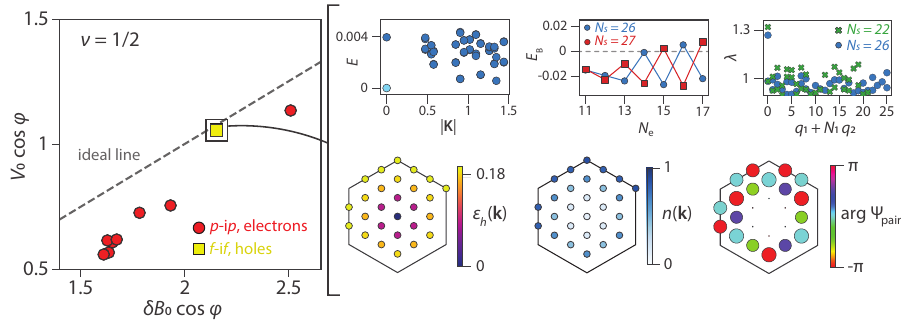}
    \caption{Evidence for $f + \im f$ hole superconductivity at $\nu = 1/2$ at location $(\delta B_0, V_0, \phi, U_0, d) =  (2.16, 1.06,0.03,0.55,0.59)$.
    Many-body spectrum was computed at system size $N_s = 26$ and number of electrons $N_e = 13$;
    2RDM spectra were computed at $N_s = 22, N_e = 11$ and $N_s = 26, N_e = 13$; momentum occupation and pair condensate wavefunction were computed at $N_s = 27, N_e = 14$.
    }
    \label{fig:other_scs_one_half}
\end{figure}

\cref{fig:other_scs_two_thirds} and \cref{fig:other_scs_one_half} show numerical evidence for chiral superconductivity for other types of superconductors found at fillings $\nu = 2/3$ and $\nu = 1/2$, respectively.
One noticeable feature is that some of the points shown display much weaker evidence for superconductivity, as compared to those shown in the main text.
As explained in \cref{sec:optim_details}, this is due to how we set up the optimization workflow, with optimizations terminating as soon as the superconductor loss function becomes negative for efficiency. 
This often leads to superconductors close to the phase boundary, which then show pairing tendencies but not unambiguous evidence for superconductivity. 

To verify that this is the case, we take some of these points and run target-phase optimization once more, but now at the much larger 27-site cluster, with the superconductor loss function defined at $N_e = 17$ for the $p + \im p$ hole superconductor, and at $N_e = 16$ for the $f-\im f$ electron superconductor.
This is numerically manageable since we are likely already near a superconductor, and therefore the optimization requires very few steps.
However, running this further optimization for the $f-\im f$ hole superconductor at half filling in \cref{fig:other_scs_one_half} was not possible under our computational constraints.

\begin{figure}
    \centering
    \includegraphics[scale=1]
    {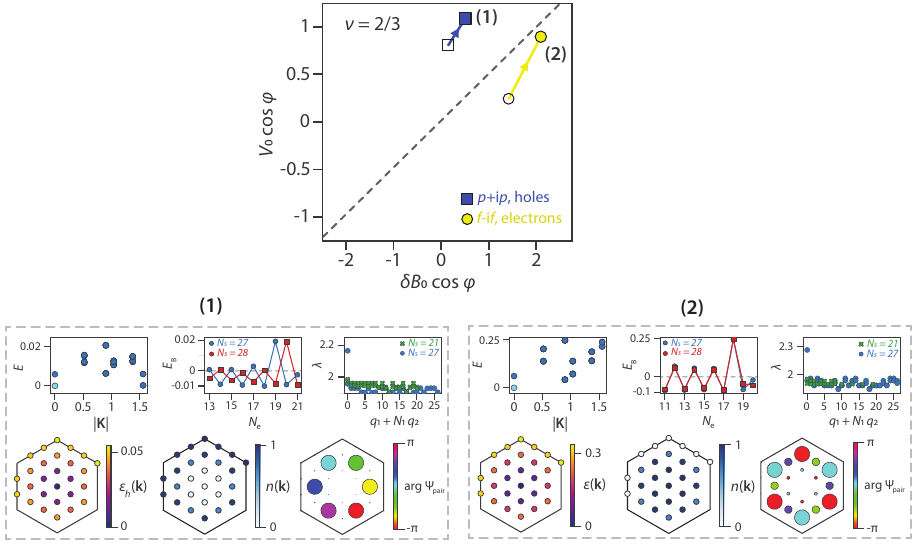}
    \caption{Optimized superconductors (filled markers), starting from previously found superconductors (empty markers) at $\nu = 2/3$.
    The locations of the optimized points in parameter space $(\delta B_0, V_0, \phi, U_0, d)$ are respectively: (1) $(0.50, 1.18,0.06, 0.85,6.82)$; (2) $(2.09, 0.90, 0.01,3.29,1.65)$.
    Calculations for the $p+\im p$ hole superconductor are carried out at the same system sizes and electron numbers as the main text.
    For the $f-\im f$ electron superconductor, many-body spectra, momentum occupation and pair condensate wavefunction were computed at system size $N_s = 27$ and number of electrons $N_e = 17$, whereas the 2RDM spectra were computed at $N_s = 21, N_e = 13$ and $N_s = 27, N_e = 17$.
    }
    \label{fig:optim_scs_two_thirds}
\end{figure}

Upon running this second optimization for the points at $\nu=2/3$, we find much clearer evidence for chiral superconductors, as shown in \cref{fig:optim_scs_two_thirds} for the $p+\im p$ hole superconductor and the $f-\im f$ electron superconductor at $\nu = 2/3$.
Notably, at these new points the two superconductors have oscillatory pair-binding energies all the way down to $\nu = 1/2$, a more thorough investigation of which we leave for future work.

\section{Single-particle quantum geometry}

The quantum geometry of a partially filled band has been shown to play a crucial role in stabilizing specific quantum phases, most prominently fractional Chern insulators.
Concretely, we first define the quantum geometric tensor~\cite{parameswaran2013fractional,liu_recent_2024}
\begin{equation}
\label{eq:qgt}
    \eta^{\mu\nu}(\mathbf{k}) = \sum_n \bra{\partial_{k_\mu}u_n(\mathbf{k})}Q(\mathbf{k})\ket{\partial_{k_\nu}u_n(\mathbf{k})}, 
\end{equation}
where $u_n(\mathbf{k})$ is a cell-periodic Bloch state in band $n$ with momentum $\mathbf{k}$, and $Q(\mathbf{k}) = 1 - \ket{u_n(\mathbf{k})}\bra{u_n(\mathbf{k})}$.
Its real and imaginary parts define the Fubini--Study metric and Berry curvature:
\begin{equation}
\label{eq:fsmetric+bc}
    g^{\mu\nu}(\mathbf{k}) = \operatorname{Re} \eta^{\mu\nu}(\mathbf{k}), \quad \Omega(\mathbf{k}) \varepsilon^{\mu\nu} = 2 \operatorname{Im} \eta^{\mu\nu}(\mathbf{k}),
\end{equation}
with $\varepsilon^{\mu\nu}$ the 2D Levi--Civita symbol. 

Geometric similarity to LLs is then quantified by two quantities, the ``trace violation'' $T$ and Berry curvature fluctuations $\sigma_\Omega$, defined as

\begin{equation}
\label{eq:spmetrics}
\begin{split}
    &T = \frac{1}{2\pi} \int_{\operatorname{BZ}} d^2k\, \operatorname{Tr}g^{\mu\nu}(\mathbf{k}) - |\Omega(\mathbf{k})|, \\   
    &\sigma_\Omega = \sqrt{\int_{\operatorname{BZ}} \frac{d^2k}{A_{BZ}}\left(\frac{\Omega(\mathbf{k})}{\Omega_0} - C\right)^2\,},
\end{split}
\end{equation}
where $A_{BZ}$ is the Brillouin zone area, $\Omega_0 = 2\pi/A_{BZ}$ and $C$ the Chern number.
In this normalization, both $T$ and $\sigma_\Omega$ are dimensionless; for the $n$-th LL, $T = 2n$ and $\sigma_\Omega=0$ ~\cite{ozawa_relations_2021}.
In particular, they both vanish for the lowest LL.

\bibliography{references}